\documentstyle[aps,multicol,epsfig,pre,psfrag,epsf]{revtex}

\begin{document}

\title{A fracture model with variable range of interaction}

\author{Raul Cruz Hidalgo$^1$, Yamir Moreno$^2$, Ferenc Kun$^3$ and  Hans J. 
Herrmann$^1$}

\address{$^1$ ICA1, University of Stuttgart, Pfaffenwaldring 27, 70569
Stuttgart, Germany.\\ $^2$ The Abdus Salam International Centre for
Theoretical Physics,\\ Condensed Matter Group, P.O. Box 586, Trieste,
I-34014, Italy.\\ $^3$ Department of Theoretical Physics, University
of Debrecen, P.\ O.\ Box: 5, H-4010 Debrecen, Hungary. }

\date{\today} 

\maketitle 

\widetext

\begin{abstract}
We introduce a fiber bundle model where the interaction among fibers
is modeled by an adjustable stress-transfer function which can
interpolate between the two limiting cases of load redistribution, the
global and the 
local load sharing schemes. By varying the range of interaction
several features of the model are numerically studied and a crossover
from mean field to short range behavior is obtained. 
The properties of the two regimes and the emergence of the crossover
in between are explored by numerically studying the dependence of the
ultimate strength of the material on the system size, the distribution
of avalanches of breakings, and of the cluster sizes of broken fibers.
Finally, we analyze
the moments of the cluster size distributions to
accurately determine the value at which the crossover is observed.

\end{abstract}
.

\begin{multicols}{2}
\narrowtext

\section{Introduction}
\label{intro}

Fracture processes have attracted the attention of the scientific
community since many years. 
Processes involving heterogeneous systems, for which
a definite and complete physical description has not been found
despite the many partial successes of the last decades, are of special
theoretical and practical interest.
\cite{h90,cha,sor2000}. In particular, the latest developments of
statistical mechanics have led to a deeper understanding of breakdown
phenomena in heterogeneous systems, but some fundamental questions
remain unsolved. The difficulties arise because in modeling
fracture of heterogeneous materials, one has to deal with systems
formed by many interacting constituents, each one having different
statistical properties related to some breaking characteristics of the
material, distributed randomly in space and/or time
\cite{h90,cha}. So, the complete analytical solution is in almost all
cases prohibitive and one has to solve the problem by means of
numerical simulations or to study simplified models which can be
analytically tractable (at least in some limits) in order to gain
physical insights that guide our understanding to more complex models.

The major challenge in dealing with fracture problems is to combine
the statistical evolution of damage across the entire macroscopic
system and the associated stress redistributions to accurately predict
the point of final rupture of the material. In doing this linkage, one
has to take care in order not to make major simplifications
particularly in the redistribution rule, where a great deal of the
physics of the problem is hidden. A very important class of approaches
to the fracture problem are the well-known Fiber Bundle Models (FBM),
which were introduced long time ago by Daniels \cite{daniels45} and
Coleman \cite{col57} and have been the subject of intense research
during the last several years
\cite{zape97,prlus00,pho83,new94,new95,vaz99,kun00,sor89,sor92,curtin98,mahesh99,hansen94,hemmer92,harlow85,klo97}.
FBM's are constructed so that
a set of fibers is arranged in parallel each one having
a statistically distributed strength. The specimen is loaded
parallel to the 
fiber direction and the fibers break if the load acting on them
exceeds their threshold value. Once the fibers begin to fail one can
choose among several load transfer rules. The simplest case is to
assume global load sharing (GLS) which means that after each fiber
failure, its load is equally redistributed among all the intact fibers
remaining in the set. This model, known as global fiber bundle model,
is a mean field approximation where long range interactions
among the elements of the system are assumed and can be solved
analytically \cite{sor92,hemmer92}. At the other extreme, one finds
the local load sharing (LLS) fiber bundle model where the load borne
by failing elements is transferred to their nearest neighbors. This
case represents short range interactions among the fibers. Other
schemes have been proposed with a relative success in describing
rupture processes at large scale like earthquakes \cite{turcotte}.

Despite their simplicity, FBM are very important because they capture
most of the main aspects of material damage and breakdown. They have
provided a deeper understanding of fracture processes and have served
as a starting point for more complex models of fiber reinforced
composites and other micro-mechanical
models\cite{mou94,har78,smi81}. However, stress redistribution in
actual heterogeneous materials should fall somewhere in between LLS
and GLS since there is an important fraction of stress redistributed
to other intact elements not localized in the neighborhood of the
failed ones, nevertheless maintaining stress concentrations around the broken
fibers. With this aim, several studies have been carried out during
the last two decades and Monte Carlo simulations have been used to
numerically study the distribution of composite strengths in 2D and 3D
for different fiber arrangements
\cite{mahesh99,bax95,bey97a,bey97b,ibn97a,ibn97b}. Nevertheless, in
order to obtain reliable conclusions the number of fibers forming the
system has to be very large which makes the numerical problem, in many
cases, too time consuming as to perform the study in a reasonable
amount of time.

In this paper, we introduce a fiber bundle model where the interaction
among fibers is modeled by an adjustable stress-transfer function,
which interpolates between the two limiting cases of load
redistribution, the global and the local load sharing schemes. By
varying the effective range of 
interaction one observes a crossover from mean field to short range
behavior. To explore the properties of the two regimes and the
emergence of the crossover in between, a comprehensive numerical study
of the model is performed. We study the dependence of the ultimate
strength of the 
material on the system size and found that the system has only one
nonzero critical load in the thermodynamic limit. When no critical
point exits, the ultimate strength of the material goes to zero
exactly as in local load sharing models as $\sim\frac{1}{ln(N)}$, with
increasing system size $N$. We also
study the distribution of avalanches of fiber breaks, and that of the
cluster sizes of broken fibers for the two 
distinct regimes and perform a moment analysis to accurately
determine the crossover value. 

The rest of the paper is organized as follows. The next section is
devoted to present the model and the way in which numerical
simulations are carried out. The results obtained are presented and
analyzed in Section \ref{section3}. The final section is devoted to
discussions and to state our conclusions.

\section{The Model}
\label{themodel}

The fracture of heterogeneous systems is characterized by the highly
localized concentration of stresses at the crack tips that makes 
possible the nucleation of new cracks at these regions such that the
actual crack grows leading to the final collapse of the system. In
elastic  materials, the stress redistribution follows a power law, 
\begin{equation}
\sigma_{add}\sim r^{-\gamma}, 
\label{eq1}
\end{equation}
where $\sigma_{add}$ is the stress increase on a material element at a
distance $r$ from the crack tip. The above general relation covers the
cases of global and local load sharing, widely used in fiber bundle
models of fracture, as the limiting cases $\gamma \rightarrow 0$, and $\gamma
\rightarrow \infty$, respectively.

Motivated by the above result of fracture mechanics we introduce a
fiber bundle model where the load sharing rule takes the form of Eq.\
(\ref{eq1}). Suppose a set of $N$ parallel fibers each one having
statistically 
distributed strength taken from a probability distribution function
$P$ 
and identified by an integer $i$, $1\leq i\leq N$. In materials
science, the Weibull distribution has been proved to be a good
empirical statistical distribution for representing fiber strength,
\[
P(\sigma)=1-e^{-(\frac{\sigma}{\sigma_{0}})^{\rho}},
\]
where $\rho$ is the so-called Weibull index, which controls the degree
of threshold disorder in the system (the bigger the Weibull index, the
narrower the range of threshold values), and $\sigma_{0}$ is a
reference load which acts as unity. Thus, to each fiber $i$ a
random threshold value $\sigma_{i_{th}}$ is assigned. The system is
driven by increasing quasistatically the load on it, which is
performed by locating the fiber which minimizes  $\sigma_i-\sigma_{i_{th}}$
and adding this amount of load to all the intact fibers
in the system. This provokes the failure of at least one fiber which
transfers its load to the surviving elements of the set. This may
provoke other fractures in the system which in turn induce tertiary
ruptures and so on until the system fails or reaches an equilibrium state
where the load on the intact fibers is lower than their individual
strengths. In this later case, the slow external driving is
applied again and the process is repeated up to the macroscopic
failure of the material. The number of broken fibers between two
successive external drivings is the size of an avalanche $s$, and the
number of parallel updatings of the lattice during an avalanche is
called its lifetime $T$.

\begin{figure}[b]
\begin{center}
\psfrag{aa}{\Large{ $ R_{max}$}}
\psfrag{bb}{\Large{ $r_{ij}$}}
\psfrag{cc}{\Large L}
\epsfig{bbllx=0,bblly=0,bburx=200,bbury=225,file=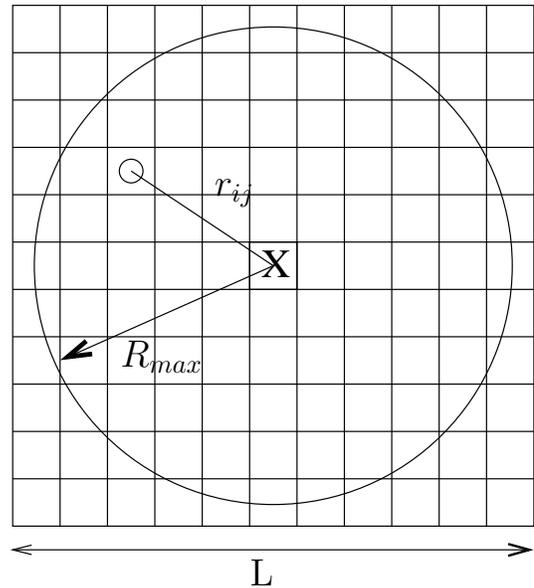,
  width=7cm}
\caption{Illustration of the model construction. {\Large{ $\times$ }}
  indicates a 
  fiber, which is going to break, and {\Large{ $\circ$}} is an intact fiber in
  the square lattice.}
\label{fig:model}
\end{center}
\end{figure}

We now focus on the load transfer process following fiber failures. We
suppose that, in general, all intact fibers have a nonzero probability
of being affected by the ongoing failure event, and that the
additional load received by an intact fiber $i$ depends on its
distance $r_{ij}$ from fiber
$j$ which has just been broken. Furthermore, elastic
interaction is assumed between fibers such that the load received by a
fiber follows the power 
law form of Eq.\ (\ref{eq1}). Hence, in our discrete model the
stress-transfer function $F(r_{ij},\gamma)$ takes the form
\begin{equation}
F(r_{ij},d)=Zr_{ij}^{-\gamma},
\label{eq4}
\end{equation}
where $\gamma$ is our adjustable parameter, $Z$ is given by the
normalization condition $Z=(\sum\limits_{i\in
I}r_{ij}^{-\gamma})^{-1}$ (the sum runs over the set $I$ of all intact
elements) and $r_{ij}$ is the distance of fiber $i$ to the rupture
point $(x_j,y_j)$, {\em i.e.}, $r_{ij}=\sqrt{(x_i-x_j)^2+(y_i-y_j)^2}$
in 2D.  Periodic boundary conditions are assumed so that the largest
$r$ value is $R_{max}=\frac{\sqrt{2}(L-1)}{2}$, where $L$ is the
linear size of the system. We note here that the assumption of
periodic boundary conditions is for simplicity. In principle, an Ewald
summation procedure would be more accurate. The model construction is
illustrated in Fig.\ \ref{fig:model}.  It is easy to see that in the
limits $\gamma\rightarrow 0$ and $\gamma\rightarrow\infty$ we recover
the two extreme cases of load redistribution in fiber bundle models:
the global load sharing and the local load sharing, respectively. We
should note here that, strictly speaking, for all $\gamma$ different
from the two limits above, the range of interaction covers the whole
lattice. However, when changing this exponent, one moves from a very
localized {\em effective} range of interaction to a truly global one
as $\gamma$ approaches zero. So, we will refer henceforth to a change
in the {\em effective} range of interaction.

In summary, during an avalanche of failure events, an intact fiber $i$ receives
at each time step $\tau$ the load borne by failing elements
$j$. Consequently, its load increases by an amount,
\begin{equation}
\sigma_i(t+\tau)=\sigma_i(t+\tau-1)+\sum\limits_{j\in
  B(\tau)}\sigma_{j}(t+\tau-1)F(r_{ij},\gamma), 
\label{eq5}
\end{equation}
where the sum runs over the set $B(\tau)$ of elements that have failed
in a time step $\tau$. Thus,
$\sigma_i(t_0+T)=\sum\limits_{\tau=1}^{T}\sigma_i(t_0+\tau)$ is the
total load element $i$ receives during an avalanche initiated at $t_0$
and which ended at $t_0+T$. In this way, when an avalanche ends, the
external field is applied again and another avalanche is
initiated. The process is repeated until no intact elements remain in
the system and the ultimate strength of the material $\sigma_c$, is
defined as the maximum load the system can support before its complete
breakdown.

Unfortunately, the complete analytical approach to the general model
introduced here is inaccessible. There are a few cases where this task
can be achieved such as the global load sharing model where the load
acting on surviving elements for a given external force $F$ is known
\cite{sor92,hemmer92} and some 1D models \cite{wu00,wu99,pho00}(which
are irrelevant for practical purposes). The main
difficulty is that in order to analytically solve the problem, one
needs to know  the transition probabilities for all the
possible paths leading the system from the state in which all the
elements are intact to the state in which they have failed. This
calculation eventually becomes impossible for large system sizes. So,
a first step is to learn from Monte Carlo simulations which,
furthermore, allows us to better understand the physical mechanisms of
fracture and to study models difficult to handle analytically as
well as to guide our search for analytical calculations.

\section{Monte Carlo simulation of the failure process}
\label{section3}

We have carried out large scale numerical simulations of the model
described above in two dimensions. The fibers are identified with the
sites of a square lattice of linear size $L$ with periodic boundary
conditions. The failure process is then simulated by varying the
effective range
of interaction between fibers by controlling $\gamma$, and recording the
avalanche size 
distribution, the cluster size distribution and the ultimate strength
of the material for several system sizes. Each numerical simulation
was performed  over at least 50 different realizations of the disorder
distribution.

\begin{figure}[b]
\begin{center}
\epsfig{file=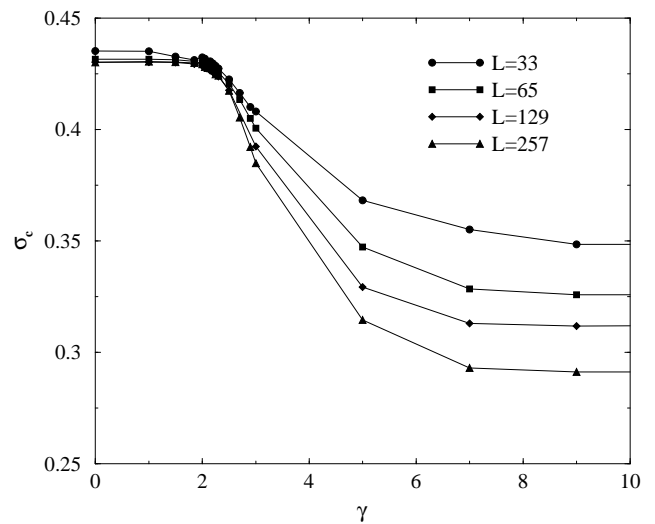,width=7.0cm,angle=-90,clip=1}
\end{center}
\caption{Ultimate strength of the material for different system sizes
as a function of the effective range of interaction $\gamma$. A crossover from mean
field to short range behavior is clearly observed.}
\label{figure1}
\end{figure}

\begin{figure}[b]
\begin{center}
\epsfig{file=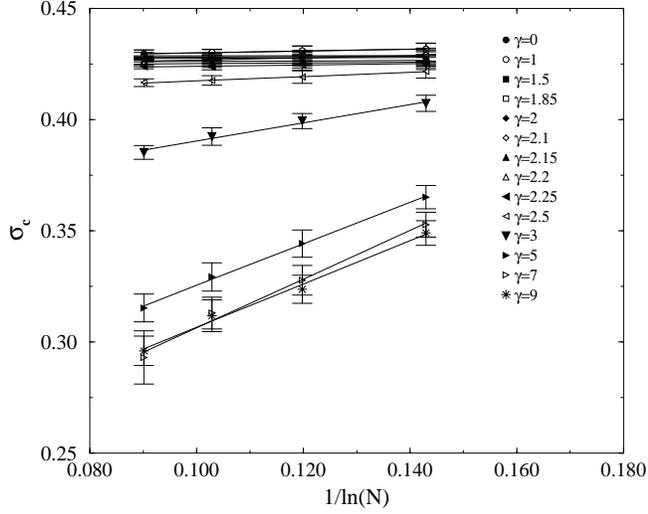,width=7.0cm,angle=-90,clip=1}
\end{center}
\caption{Variation of the material strength with $N$  for several
values of $\gamma$. Note that when $\gamma$ increases the critical load
vanishes in the thermodynamic limit, whereas, for small $\gamma$ it has a
nonzero value independent on the system size.}
\label{figure2}
\end{figure}

Figure\ \ref{figure1} shows the ultimate strength of the material for
different values of the parameter $\gamma$ and several system sizes from
$L=33$ to $L=257$. Clearly, two distinct regions can be
distinguished. For small $\gamma$, $\sigma_c$ is independent, within
statistical errors, of both the effective range of interaction and the
system size. At a given point $\gamma=\gamma_c$ a crossover is
observed, where $\gamma_c$ falls in the vicinity of $\gamma=2$. The 
region $\gamma>\gamma_c$ might eventually be 
further divided into two parts, the first region
characterized by the dependence of the ultimate strength of the bundle
on both the system size and the effective range of interaction; and a second
region where $\sigma_c$ only depends on the system size. This would mean 
that there might be two transition points in the model, for which the
system displays qualitatively and quantitatively different
behaviors. For $\gamma\leq \gamma_{c}$ the ultimate strength of the bundle
behaves as in the limiting case of global load sharing, whereas for
$\gamma\geq \gamma_{c}$ the local load sharing
behavior seems to prevail. Nevertheless, the most important feature is
that when 
decreasing the effective range of interaction in
the thermodynamic limit, for $\gamma>\gamma_{c}$, the critical load is
zero. This observation is further supported by Fig.\ \ref{figure2}, where we
have plotted the evolution of $\sigma_c$ as a function of
$\frac{1}{\ln N}$ for different values of the exponent $\gamma$. Here, the
two limiting cases are again clearly differentiated. For large $\gamma$ all
curves decreases when $N\rightarrow \infty$ as
\begin{equation}
\sigma_c(N)\sim \frac{\alpha}{\ln N}
\label{eq6}
\end{equation}
This qualifies for a genuine short range
behavior as found in LLS models where the same relation was obtained 
for the asymptotic strength of the bundle \cite{harlow85,klo97}. 
It is worth noting that in the model we are
analyzing, the limiting case of local load sharing corresponds to
models in which short range interactions are considered to affect the
nearest and the next-nearest neighbors. In the transition region, the
maximum load the system can support also decreases as we approach
the thermodynamic limit, but in this case much slower than for $\gamma\gg
\gamma_{c}$. It has been pointed out that for some modalities of stress
transfer, which can be considered as intermediate between GLS
and LLS, $\sigma_c$ decreases for large system sizes following the
relation $\sigma_c\sim \frac{1}{\ln(\ln N)}$ as in the case of
hierarchical load transfer models \cite{new91}. In our case,
we have fitted our results with this relation but we have not obtained
a single collapsed curve because the slopes continuously vary until the
LLS limit is reached. Finally, the region where the ultimate stress
does not depend on the system size shows the behavior expected for the
standard GLS model, where the critical load can be exactly computed as
$\sigma_c=(\rho e)^{-1/\rho}$ for the Weibull distribution. The
numerical values obtained for $\rho=2$ are in good agreement 
with this later expression.

\begin{figure}[b]
\begin{center}
\epsfig{file=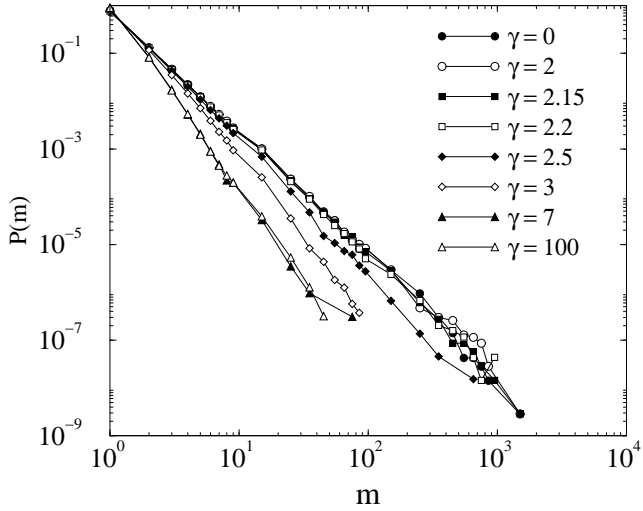,width=8.5cm,angle=0,clip=1}
\end{center}
\caption{Avalanche size distributions for different values of the
exponent of the stress-transfer function $\gamma$. The upper group of
curves can be very well fitted with a straight line with a slope
$\tau=-\frac{5}{2}$ ($L=257$).}
\label{figure3}
\end{figure}

The fracture process can also be investigated by looking at the
recursory activity before the complete breakdown. The statistical
properties of rupture sequences are characterized by the avalanche
size distribution which from the experimental point of view could be
related to the acoustic emissions generated during the fracture of
materials \cite{gar97,gar98,maes98,pet94}. Figure\ \ref{figure3} shows
the avalanche size distribution for different values of
$\gamma$. Again, we observe that for decreasing effective range of
interaction (increasing $\gamma$) there is a crossover in the
distribution of avalanche sizes. The upper curves can be very well
fitted by a power law $P(m)\sim m^{-\tau}$, with
$\tau\approx\frac{5}{2}$, the value obtained for long range
interactions \cite{hansen94,hemmer92,harlow85,klo97}.

As soon as the localized nature of the interaction becomes dominant
$\gamma>\gamma_c$, the power law dependence of the avalanche size
distribution with the exponent $\tau\approx\frac{5}{2}$ does not apply
anymore.  The lack of a characteristic size is a fingerprint of a
highly fluctuating activity that could be related to the very nature
of the long range interactions. The avalanche size distribution is a
measure of causally connected broken sites and the spatial
correlations in this limit are ruled out. All the intact elements have
a nonzero chance to fail independently of the (spatial) rupture
history, and any given element could be near to its rupture point
regardless of its position in the lattice.  This is not indeed the
case when $\gamma$ is large enough and the short range interaction
prevails. Now, the spatial correlations are important and
concentration of stress takes place in the fibers located at the
perimeter of an already formed cluster. Fibers far away the clusters
of broken elements have significantly lower stresses and thus the size
of the largest avalanche is reduced as well as the number of failed
fibers belonging to the same avalanche, leading to a lower precursory
activity.

\begin{figure}[b]
\begin{center}
\epsfig{file=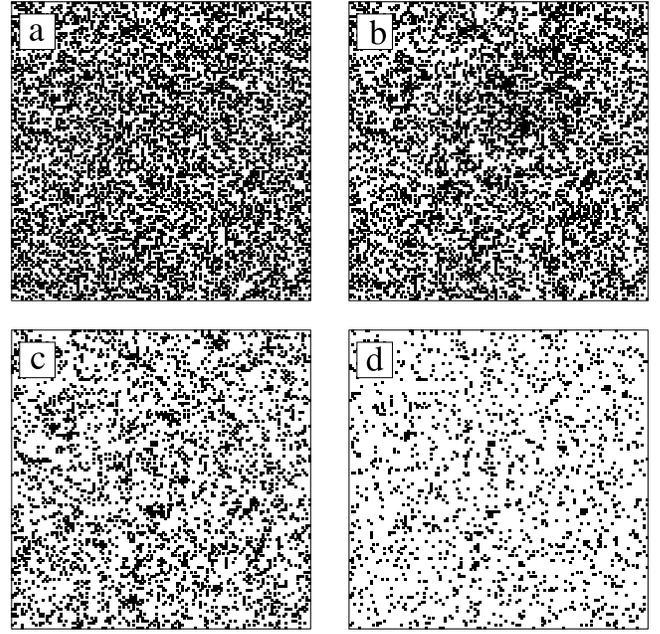,width=8.5cm,angle=0,clip=1}
\end{center}
\caption{Snapshots of the clusters just before the complete breakdown
of the material. The change in the structure of the clusters can be
seen. The values of $\gamma$ are: {\em a)} $\gamma=0$, {\em
b)} $\gamma_c=2.2$, {\em c)} $\gamma=3$, and {\em d)} $\gamma=9$. }
\label{figure4}
\end{figure}

\begin{figure}[b]
\begin{center}
\epsfig{file=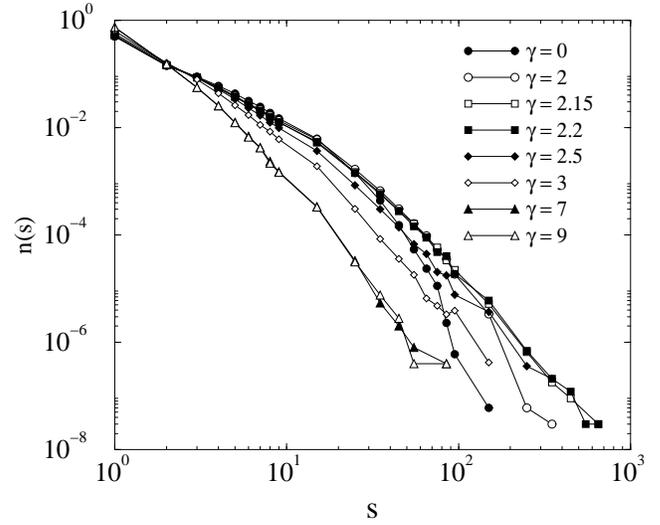,width=8.5cm,angle=0,clip=1}
\end{center}
\caption{Cluster size distributions for different values of the 
stress-transfer function exponent $\gamma$. Clearly, two different groups 
of curves can be  distinguished as found for other quantities 
also ($L=257$).}
\label{figure5}
\end{figure}

A further characterization of what is going on in the fracture process
can be carried out by focusing on the properties of clusters of broken
fibers. The clusters formed during the evolution of the fracture
process are sets of spatially connected broken sites on the square
lattice \cite{zape97,kun00}. We consider the clusters just before the
global failure and they are defined taking into account solely nearest
neighbors connections. It is important to note that the case of global
load sharing does not assume any spatial structure of fibers since it
corresponds to the mean field approach. However, in our case it is
obtained as a limiting case of a local load sharing model on a square
lattice, which justifies the cluster analysis also for GLS.  Fig.\
\ref{figure4} illustrates how the cluster structure just before
complete breakdown changes for various values of $\gamma$.  We have
also recorded the cluster size distribution as a function of the
effective range of interaction. Figure\ \ref{figure5} shows the size
distribution $n(s_c)$, of the two-dimensional clusters for several
values of the exponent of the load sharing function. The distributions
have clearly two groups as found for other quantities also.

In the limit where the long range interaction dominates, the clusters
are randomly distributed on the lattice indicating that there is no
correlated crack growth in the system as well as that the stress is
not concentrated in regions. The cluster structure of the limiting
case of $\gamma=0$ can be mapped to percolation clusters on a square
lattice generated with the probability $0 < P(\sigma_c) < 1$, where
$\sigma_c$ is the fracture strength of the fiber bundle. However, the
value of $P(\sigma_c)$ depends on the Weibull index $\rho$ and is
normally different from the critical percolation probability
$p_c=0.592746$ of the square lattice. Equality $P(\sigma_c) = p_c$ is
obtained for $\rho=1.1132$, hence, for physically relevant $\rho$
values used in simulations the system is below $p_c$ at complete
breakdown. This argument also justifies the exponential-like shape of
the cluster size distributions of GLS in Fig.\ \ref{figure5}.  This
picture radically changes when the short range interaction
prevails. In this case, the stress transfer is limited to a
neighborhood of the failed elements and there appear regions where a
few isolated cracks drive the rupture of the material by growth and
coalescence. Thus, the probability of the existence of a weak region
somewhere in the system is high and a weak region in the bundle may be
responsible for the failure of the material. The differences in the
structure of clusters also explain the lack of a critical strength
when $N$ goes to infinity in models with local rearrangement of
stress. Since in the GLS model the clusters are randomly dispersed
across the entire lattice, the system can ``store'' more damage or
stress, whereas for LLS models a small increment of the external field
may provoke a run away event ending with the macroscopic breakdown of
the material.

\begin{figure}[t]
\begin{center}
\epsfig{file=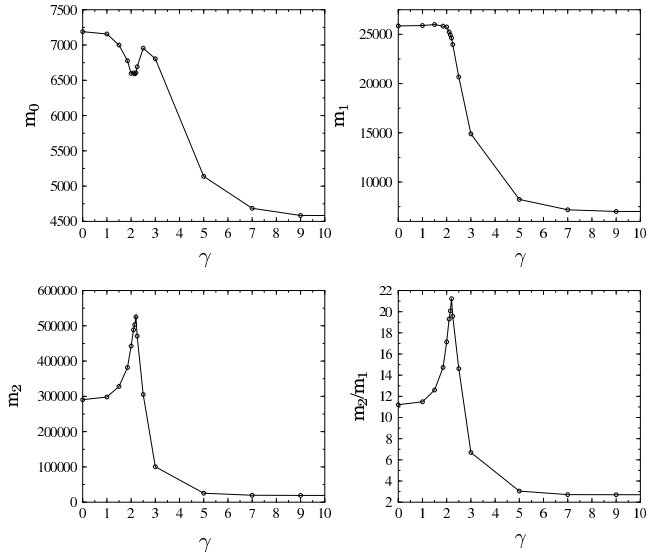,width=8.5cm}
\end{center}
\caption{Moments of the cluster size distribution as a function of $\gamma$
(see text for details on the definition of $m_k$). A sharp maximum is
observed at $\gamma=\gamma_c\sim2.2$ for the average cluster size $\langle s_c 
\rangle=\frac{m_2}{m_1}$.}
\label{figure6}
\end{figure}

Up to now, the change of the behavior of the system was observed for
a certain value of $\gamma$ analysing various measured
quantities. All these numerical results suggest that the crossover
between the two regimes occurs
in the vicinity of $\gamma =2$. Further support for the precise value
of $\gamma_c$ can be obtained by studying the change in the cluster
structure of broken fibers. The moments of $n(s_c)$ 
defined as 
\begin{equation}
m_k\equiv\int s_c^k n(s_c) ds
\label{eq7}
\end{equation}
where $m_k$ is the $k$th moment, describe much of the physics
associated with the breakdown process. We will use these moments
to quantitatively characterize the point where the crossover from mean
field to short range behavior takes place. The zero moment $n_c=m_o$
is the total number of clusters in the system and is plotted in Fig.\
\ref{figure6}a as a function of the parameter $\gamma$. Figure\
\ref{figure6}b represents the variation of the total number of broken
sites $N_c$ (the first moment $N_c=m_1$) when $\gamma$ increases. It turns
out that up to a certain value of the effective range of interaction, $N_c$
remains constant and then it decreases fast until a second plateau
seems to arise. Note that the constant value of $N_c$ for small
$\gamma$ is in agreement with the value of the fraction of broken fibers
just before the breakdown of the material in mean field models. This
property clearly indicates a change in the evolution of the failure
process and may serve as a criterion to calculate the crossover
point. However, a more abrupt change is observed in the average
cluster size $\langle s_c \rangle$ at varying $\gamma$. According to the
moments description, the average cluster size is equal to the second
moment of the cluster distribution divided by the total number of
broken sites, {\em i.e.} $\langle s_c \rangle=m_2/m_1$. It can be
seen in Fig.\ \ref{figure6}d that $\langle s_c \rangle$ has a sharp
maximum at $\gamma=2.2\pm0.1$, and thus the average cluster size 
drastically changes at this point, which again suggests the crossover
point to be in the vicinity of $\gamma_c=2$.

\begin{figure}[b]
\begin{center}
\epsfig{file=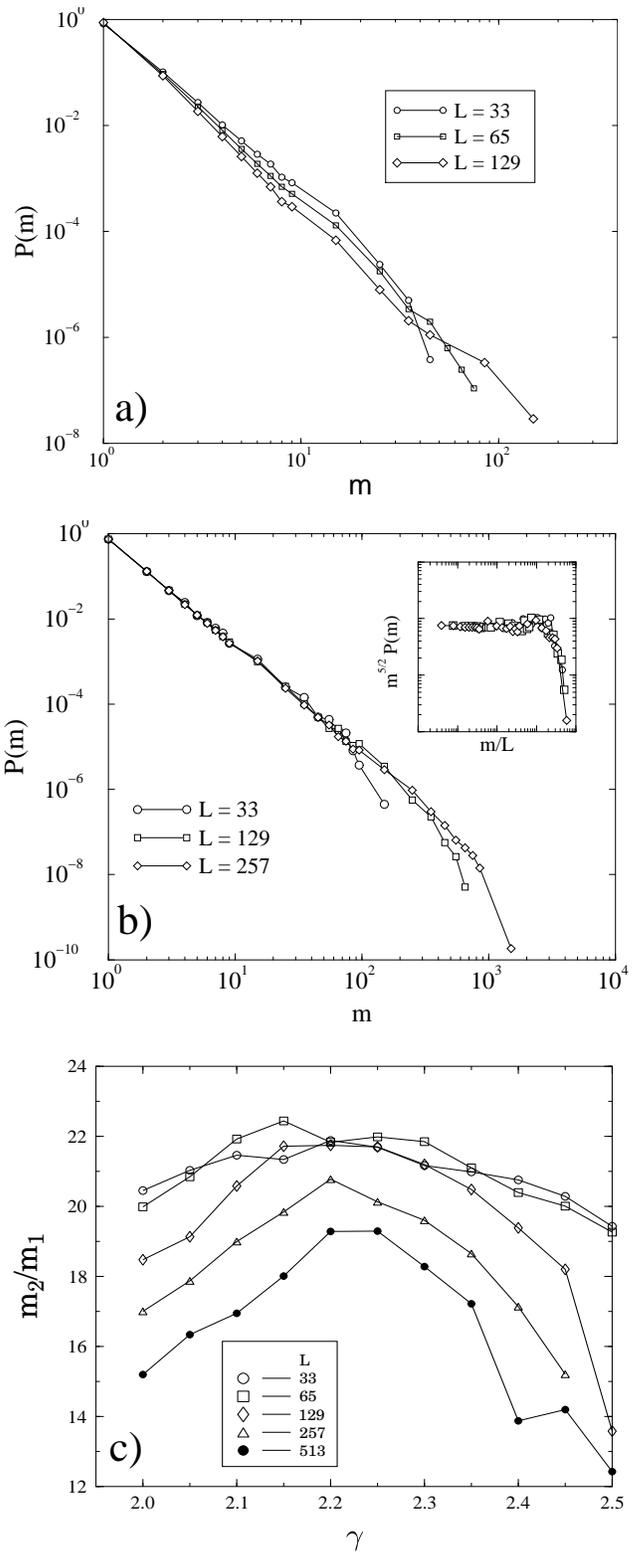,width=8.5cm,angle=0,clip=1}
\end{center}
\caption{Finite size scaling analysis. $a)$ Scaling of the cutoff with
the system size for the local load sharing case, $b)$ Scaling of the
cutoff with the system size for the global load sharing case, and $c)$
Average cluster size, $\langle s_c \rangle=\frac{m_2}{m_1}$, for
different system sizes. Note that in $c)$ the position of $\gamma_c$
does not change.}
\label{figure8}
\end{figure}

We now discuss the finite size scaling (FSS) of the avalanche
distributions. For local load sharing one expects that the cutoff in
the avalanche distribution does not scale with $L$ while for global
load sharing the cutoff should scale with the size of the system. We
have plotted in Fig.\ \ref{figure8} the avalanche size distribution
for several system sizes. As it can be observed, the FSS hypothesis is
verified for the values of the exponent $\gamma$ corresponding to the
global (Fig.\ \ref{figure8}$b$) and the local (Fig.\ \ref{figure8}$a$)
load sharing cases. Figure \
\ref{figure8}$c$ shows the moment analysis for five different system
sizes in the range $2.0 \leq \gamma \leq 2.5$. It can be seen that the
position of the maximum of the $m_2/m_1$ curves is always at $\gamma = 2.2
\pm 0.1$, it does not scale with the system size. To determine the
position of the crossover point more accurately we also analyzed the
behaviour of $\alpha$ characterizing the strength of logarithmic size
effect in Eq.\  (\ref{eq6}), as a function of $\gamma$. From these
studies it turned out that consistent interpretation of the numerical
results can be given assuming that the crossover occurs in the
vicinity of $\gamma_c = 2.0$ but stonger statement cannot be drawn due
to the limited precision of calculations.


\section{Discussion and Conclusions}
\label{section4}

In the limiting case of global load sharing the breaking of fibers is
a completely random nucleation process, there is no correlated crack
growth in the system, and the fiber failure which results in the
catastrophic avalanche occurs at a random position in the
system. As long as this microscopic damage mechanism holds when changing the
exponent $\gamma$, globally the system will behave in a global load sharing
manner. On the other hand, when the load sharing is very localized, at
the beginning of the failure process we get random nucleation of
microcracks but later, correlated growth of clusters of broken fibers
occurs. It then follows that along the perimeter of the clusters there
is a high stress concentration and the final avalanche is driven by a
fiber located at the perimeter of one of the clusters (the dominant
one). At the fibers far away from the perimeter, the stress
concentration is significantly lower, and the stress distribution is
very inhomogeneous. In the case of localized load sharing this mechanism 
gives rise also to the logarithmic size effect as obtained also for the 
random fuse model \cite{kahng88}.

An interesting aspect to be explored in future work is whether there
is a second transition point in the model when the $\gamma$-dependence
of $\sigma_c$ seems to disappear or it is just a crossover. In the
case of localized load sharing the global failure is caused, as noted
before, by the instability of a cluster of broken fibers, defining a
critical cluster size. Furthermore, the exact value of $\gamma_c$
might depend on the amount of disorder in the system which will also
be subject of future studies. Preliminary studies of this issue
indicate that the transition value $\gamma_c$ gets slightly smaller as
the system becomes more homogeneous (increasing $\rho$).

Before setting our conclusions, we would like to remark that a similar 
transition to the one obtained here has been observed in other models. It is
well-known that long range interactions between spins can affect the
critical behavior of magnets. If we consider the addition of a term
$\sum_{r,r'}V(r-r')s(r)s(r')$ to an Ising Hamiltonian, where $V(r)\sim
r^{-(d+\sigma)}$ corresponds to a long range interaction decaying as a
power law, a crossover to long range behavior is obtained provided
that $\sigma<2-\eta_{sr}$, where $\eta_{sr}$ is the value of the
critical exponent $\eta$ for short range
force\cite{car96,sak77,bre76}. A similar behavior has also been
reported in percolation phenomena with long range correlations
\cite{sahimi98}.

In summary, we have studied a fracture model of the fiber bundle type
where the interaction among fibers is considered to decay as a power
law of the distance from an intact element to the rupture point. Two
very different regimes are found as the exponent of the
stress-transfer function varies and a crossover point is identified at
$\gamma=\gamma_c$. The strength of the material for $\gamma<\gamma_c$
does not depend on both the system size and $\gamma$ qualifying for
mean-field behavior, whereas for the short range regime, the critical
load vanishes in the thermodynamic limit. The behavior of the model at
both sides of the crossover point was numerically studied by recording
the avalanche and the cluster size distributions. The numerical
results suggest that the crossover point
falls in the vicinity of $\gamma_c=2.0$. Finally, we have outlined
some general ideas which will be the subject of a forthcoming
publication. 

\section{acknowledgments}
\label{final}

Y.\ M.\ thanks A.\ F.\ Pacheco and A.\ Vespignani for useful comments
on the manuscript. This work began while one of the authors (Y.\ M.\ )
was visiting the Institute for Computational Physics (ICA1) at the
University of Stuttgart. Its financial support and hospitality are
gratefully acknowledged. F.\ K.\ acknowledges financial support of the
B\'olyai J\'anos Fellowship of the Hungarian Academy of Sciences and
of the Research Contract FKFP 0118/2001. This work was supported by
the project SFB381, and by the NATO grant PST.CLG.977311.

\end{multicols}
\end{document}